\documentclass{article}
\usepackage{ijcai23}

\usepackage{times}
\usepackage{soul}
\usepackage{url}
\usepackage[hidelinks]{hyperref}
\usepackage[utf8]{inputenc}
\usepackage[small]{caption}
\usepackage{graphicx}
\usepackage{amsmath}
\usepackage{amsfonts}
\usepackage{amsthm}
\usepackage{booktabs}
\usepackage{algorithm}
\usepackage{algorithmic}
\usepackage[switch]{lineno}
\usepackage{multirow}
\usepackage{marvosym}

\title{SemiGNN-PPI: Self-Ensembling Multi-Graph Neural Network for Efficient and Generalizable Protein-Protein Interaction Prediction}

\author{
Ziyuan Zhao$^{1,2,}$\thanks{equal contribution}
\and
Peisheng Qian$^{1,*}$\and
Xulei Yang$^{1, {\textrm{\Letter}}}$\and
Zeng Zeng$^3$\and
Cuntai Guan$^2$\and
\\Wai Leong Tam$^4$\And
Xiaoli Li$^{1,2}$
\affiliations
$^1$Institute for Infocomm Research (I$^2$R), A*STAR, Singapore\\
$^2$School of Computer Science and Engineering (SCSE), Nanyang Technological University, Singapore\\
$^3$School of Microelectronics, Shanghai University, China\\
$^4$Genome Institute of Singapore (GIS), A*STAR, Singapore
\emails
\{zhaoz, qianp ,yangx\}@i2r.a-star.edu.sg,
zengz@shu.edu.cn,
ctguan@ntu.edu.sg,
tamwl@gis.a-star.edu.sg,
xlli@i2r.a-star.edu.sg
}

\begin{document}

\maketitle

% Multidisciplinary Topics and Applications

\begin{abstract}

% The abstract should be no more than 200 words long

% 1. multi-graph
% 2. mean teacher
% 3. graph alignment
Protein-protein interactions (PPIs) are crucial in various biological processes and their study has significant implications for drug development and disease diagnosis. 
Existing deep learning methods suffer from significant performance degradation under complex real-world scenarios due to various factors,~\emph{e.g.}, label scarcity and domain shift. 
In this paper, we propose a self-ensembling multi-graph neural network (SemiGNN-PPI) that can effectively predict PPIs while being both efficient and generalizable. 
In SemiGNN-PPI, we not only model the protein correlations but explore the label dependencies by constructing and processing multiple graphs from the perspectives of both features and labels in the graph learning process. 
We further marry GNN with Mean Teacher to effectively leverage unlabeled graph-structured PPI data for self-ensemble graph learning. We also design multiple graph consistency constraints to align the student and teacher graphs in the feature embedding space, enabling the student model to better learn from the teacher model by incorporating more relationships. 
Extensive experiments on PPI datasets of different scales with different evaluation settings demonstrate that SemiGNN-PPI outperforms state-of-the-art PPI prediction methods, particularly in challenging scenarios such as training with limited annotations and testing on unseen data.

\end{abstract}

\section{Introduction}

% Background of PPI

Protein-protein Interactions (PPIs) are central to various cellular functions and processes, such as signal transduction, cell-cycle progression, and metabolic pathways~\cite{acuner2011transient}.
Therefore, the identification and characterization of PPIs are of great importance for understanding protein functions and disease occurrence, which can potentially facilitate therapeutic target identification~\cite{petta2016modulation} and the novel drug design~\cite{skrabanek2008computational}.
In past decades, high-throughput experimental methods,~\emph{e.g.}, yeast two-hybrid screens (Y2H)~\cite{fields1989novel}, and mass spectrometric protein complex identification (MS-PCI)~\cite{ho2002systematic} have been developed to identify PPIs.
Nevertheless, genome-scale experiments are expensive, tedious, and time-consuming while suffering from high error rates and low coverage~\cite{luo2015highly}.
As such, there is an urgent need to establish reliable computational methods to identify PPIs with high quality and accuracy.

% Introduce ML & DL methods for PPI
% Outline different computational methods developed for PPI prediction, such as machine learning, structural modeling, and network-based approaches.

In recent years, a large variety of high-throughput computational approaches for PPI prediction have been proposed, which can be broadly divided into two groups: classic machine learning~(ML)-based methods~\cite{KNN-PPI,Byes-PPI,SVM-PPI,RF-PPI,DTree-PPI} and deep learning~(DL)-based methods~\cite{sun2017sequence,DeepPPI,DPPI,PIPR,GNN-PPI}.
Compared to classic ML methods, DL algorithms are capable of processing complicated and large-scale data and extracting useful features automatically, achieving significant success in a diverse range of bioinformatics applications~\cite{min2017deep,soleymani2022protein}, including PPI prediction~\cite{soleymani2022protein}. 
Most existing DL-based methods treat interactions as independent instances, ignoring protein correlations.
PPI can be naturally formulated as graph networks with proteins and interactions represented as nodes and edges, respectively~\cite{margolin2006aracne,pio2020exploiting}. 
To improve PPI prediction performance, recent works~\cite{yang2020graph,GNN-PPI} have been proposed to investigate the correlations between PPIs using various graph neural network (GNN) architectures~\cite{kipf2016semi,xu2018how}. However, they are limited by ignoring learning label dependencies for multi-type PPI prediction. 
It has recently become common practice to employ Graph Convolutional Networks (GCNs) to capture label correlation in a wide range of multi-label tasks~\cite{chen2019multi,wang2020multi}. Nevertheless, multi-label learning utilizing label graphs predominantly works in the visual domain and has yet to be extended to PPI prediction tasks.

In general, a desired PPI prediction framework should be efficient, transferable, and generalizable, whereas two major bottlenecks deriving from imperfect datasets have hindered the development of such models.
\textbf{Label scarcity:} Despite the tremendous progress in PPI research using various computational and experimental methods, many interactions still need to be annotated from experimental data. 
Consequently, only a small portion of labeled samples can be used for model training. 
It can be a significant bottleneck in obtaining robust and accurate PPI prediction models. 
\textbf{Domain shift:} Most existing methods are only developed and validated using in-distribution data (~\emph{i.e.}, trainset-homologous testsets), receiving severe performance degradation when being deployed to unseen data with different distributions (~\emph{i.e.}, trainset-heterologous testsets). Although~\cite{GNN-PPI} design new evaluations to better reflect model generalization, giving instructive and consistent assessment across datasets, the domain shift issue still needs to be fully explored for PPI prediction. 
Therefore, how to deal with imperfect data for improving model efficiency and generalization remains a vital issue in PPI prediction. Recent studies~\cite{zhang2021semi,zhao2022uda} show that self-ensemble methods with semi-supervised learning (SSL)~\cite{tepens2017,meanteacher} have demonstrated effectiveness in addressing both label scarcity and domain shift.

In this work, to tackle the above challenges and limitations, we propose an efficient and generalizable \textbf{PPI} prediction framework, referred to as \underline{\textbf{S}}elf-\underline{\textbf{e}}nsembling \underline{\textbf{m}}ult\underline{\textbf{i}}-\underline{\textbf{G}}raph \underline{\textbf{N}}eural \underline{\textbf{N}}etwork (\textbf{SemiGNN-PPI}). 
Firstly, we propose leveraging graph structure to model protein correlations and label dependencies for multi-graph learning. 
Specifically, we learn inter-dependent classifiers to extract information from the label graph, which are then applied to the protein representations aggregated by neighbors in the protein graph for multi-type PPI prediction.
Secondly, we propose combining GNN with Mean Teacher~\cite{meanteacher}, a powerful SSL model, to explore unlabeled data for self-ensemble graph learning. In our framework, the student model learns to classify the labeled data accurately and also distills the knowledge beneath unlabeled data from the teacher model with multiple graph consistency constraints for improving the model performance under complex scenarios.
To the best of our knowledge, this is the first study to explore efficient and generalizable multi-type PPI prediction. Precisely, the main contributions of the work can be summarized as follows:

\begin{itemize}
  % \item We investigate a complex but realistic scenario in PPI prediction, where 1) labels are limited and classes are unbalanced, and 2) the train and test data are heterogeneous. We present a novel PPI prediction approach, SemiGNN-PPI, to simultaneously exploit label correlation for unbalanced classes and leverage unlabeled PPI data for label scarcity.

  \item For multi-type PPI prediction, we first investigate the limitations and challenges of existing methods under complex but realistic scenarios, and then propose an effective \underline{\textbf{S}}elf-\underline{\textbf{e}}nsembling \underline{\textbf{m}}ult\underline{\textbf{i}}-\underline{\textbf{G}}raph \underline{\textbf{N}}eural \underline{\textbf{N}}etwork-based \underline{\textbf{PPI}} prediction (\textbf{SemiGNN-PPI}) framework for improving model efficiency and generalization.
  
  % \item To model the label correlation, we construct a label GCN based on pre-trained PPI label embeddings and label co-occurrences. We further combine GNN and the Mean Teacher paradigm to utilize unlabeled data effectively. Moreover, we generate embedding graphs in the self-ensemble models and encourage graph alignment as additional consistency regularization. 
  \item In SemiGNN-PPI, we construct multiple graphs to learn correlations between proteins and label dependencies simultaneously. We further advance GNN with Mean Teacher to effectively utilize unlabeled data by consistency regularization with multiple constraints. 
  
  \item Extensive experiments on three PPI datasets with different settings demonstrate that SemiGNN-PPI outperforms other state-of-the-art methods for multi-label PPI prediction under various challenging scenarios.

  % We conduct extensive experiments on three datasets with different train-test partition schemes. We present superior results over existing state-of-the-art approaches for multi-label PPI prediction. 
\end{itemize}

\begin{figure*}[t]
    \centering
    \includegraphics[width = 0.8\textwidth]{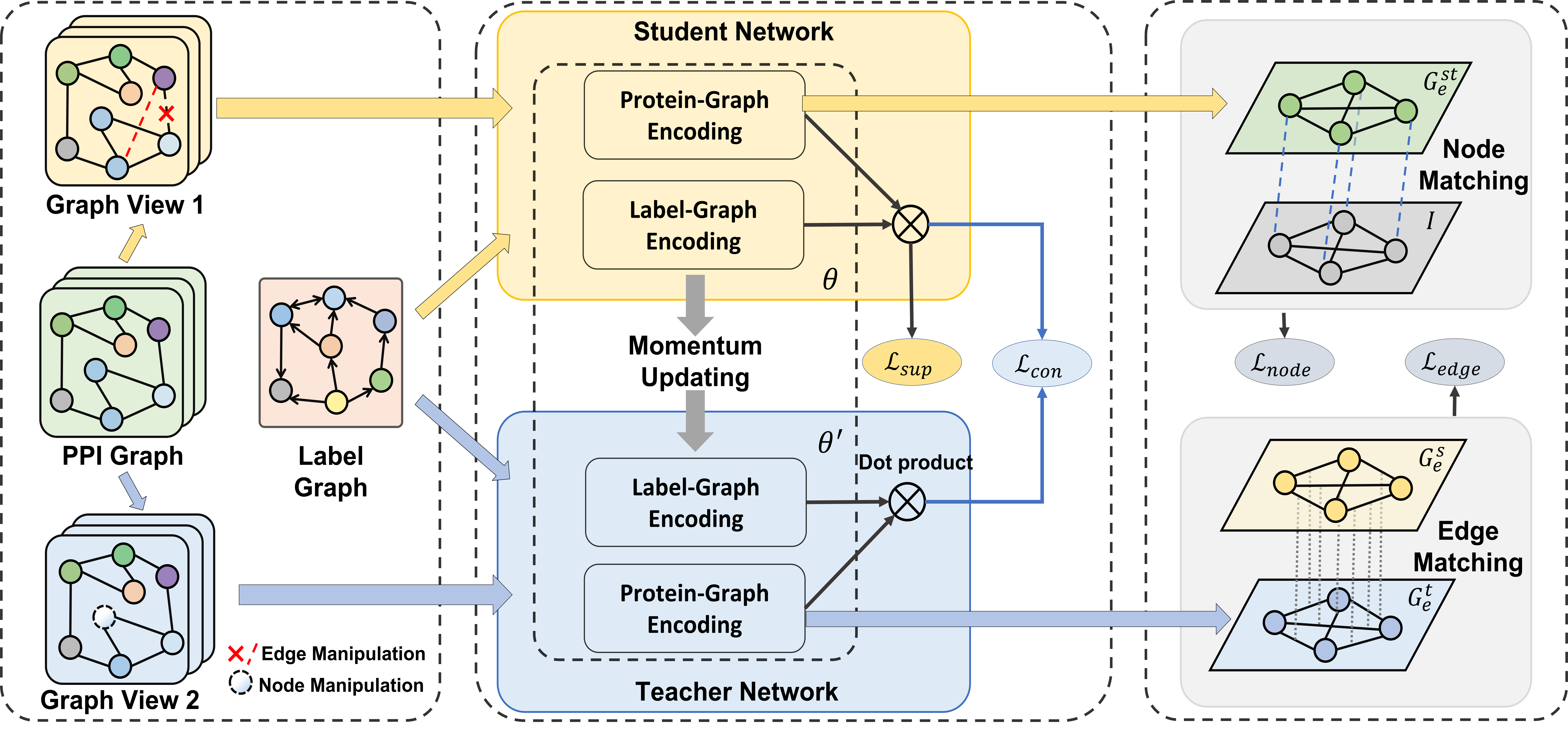}
    \caption{The overall framework of SemiGNN-PPI. First, we generate two augmented graph views with node and edge manipulations. Then, protein graphs and label graphs are fed into the multi-graph teacher-student network, which models both protein relations and label dependencies for self-ensemble learning. Simultaneously, to better capture fine-grained structural information, we align student and teacher feature embeddings by jointly optimizing multiple graph consistency constraints (node matching and edge matching).}
    \label{fig:archi}
\end{figure*}

\section{Related Work}
\noindent\textbf{Protein–Protein Interaction Prediction.} Amino acid sequence-based methods have received considerable attention in PPI prediction. Early works leverage machine learning~(ML) techniques~\cite{KNN-PPI,DTree-PPI,Byes-PPI,SVM-PPI,RF-PPI} to map pairs of handcrafted sequence features of proteins to interaction types. With the advent of deep learning~(DL), more recent works have utilized deep neural networks~\cite{sun2017sequence,DPPI,DeepPPI,PIPR,GNN-PPI} to automatically extract features from protein sequences for enhancing feature representation. Furthermore, the latest works consider protein correlations and utilize graph neural networks (GNN) to model graph-structured PPI data~\cite{yang2020graph,kipf2016semi,GNN-PPI}. However, it is essential to explore label dependencies for improving the model performance, which has long been ignored for multi-type PPI prediction. Moreover, the generalization and efficiency problems for PPI prediction are still under-explored under complex scenarios, such as data scarcity and distribution shift.

% PPI based on amino acid sequence has attracted research attention in recent years~\cite{PIPR,GNN-PPI}. Early works leverage traditional machine learning methods to map pairs of protein sequences to protein interactions, including SVM, kNN, Random Forest~\cite{wong2015detection} and multi-layer perceptron (MLP). The PPI prediction performance of existing machine learning methods is bounded by their capability in extracting protein features. Recent works apply deep learning approaches to PPI prediction for enhanced feature representation, including Convolutional Neural Networks (CNN) and Recurrent Neural Networks (RNN)~\cite{DNN-PPI,DPPI,PIPR}. Deep learning methods focus on individual protein embeddings but ignore the correlations among the PPI network. Graph Neural Networks have been proposed to exploit the PPI network, including Graph Convolutional Network (GCN)~\cite{yang2020graph,kipf2016semi} and Graph Isomorphism Network (GIN)~\cite{GNN-PPI}. Despite the performance increment, the multi-label PPI prediction problem is still under-explored in conjunction with data sparsity and distribution shift.

\noindent\textbf{Multi-Label Learning.} MLL addresses the problem of assigning multiple labels to a single instance. It has been utilized successfully in numerous fields,~\emph{e.g.}, computer vision~\cite{liu2021emerging,9897968}.  Traditional MLL methods typically train independent classifiers for all labels but fail to consider the potential label interdependence, leading to suboptimal performance. Recent trends in MLL incorporate deep learning to capture the label dependencies~\cite{wang2016cnn,guo2019breaking}. For example, CNN-RNN~\cite{wang2016cnn} leverages recurrent neural networks (RNNs) to transform the label vectors into an embedded space to learn label correlations implicitly. More recently, graph-based MLL methods have aroused great attention from researchers~\cite{chen2019multi,wang2020multi}. Especially, ML-GCN~\cite{chen2019multi} successfully applies Graph Convolutional Network (GCN) by constructing a directed graph over object labels to explicitly model the label dependencies adaptively. In this regard, we propose to explore correlations between PPI types with GCN on the structured label graph for more accurate PPI prediction.

% Multi-label learning assigns multiple labels to each sample. It is an emerging topic in various research fields,~\emph{e.g.}, protein function classification and video content tagging~\cite{liu2021emerging}. Multi-label methods can be classified into three categories: 
% 1) embedding-based methods, which projects the labels into a low dimensional space~\cite{hsu2009multi}; 2
% ) tree-based methods, which hierarchically divides the entire set of labels into smaller subsets~\cite{prabhu2014fastxml}; 
 % 3) one-vs-all methods which trains independent classifiers for all labels. Recent advances in multi-label learning involves deep learning methods to capture inter-dependencies among labels~\cite{you2019attentionxml,guo2019breaking}. 
 % Particularly, ML-GCN successfully applies Graph Convolutional Network (GCN) tn multi-label classification for images by building a directed graph over object labels to capture label correlations~\cite{chen2019multi,wang2020multi}. Inspired by ML-GCN, we construct the PPI label graph as a component of our multi-graph neural network.

\noindent\textbf{Learning from Imperfect Data.}
In recent years, deep learning has made tremendous progress in numerous domains,~\emph{e.g.}, computer vision and bioinformatics. However, the applicability of deep learning is limited by heavy reliance on training data. We rarely have a perfect dataset for model training~\cite{tajbakhsh2020embracing,bekker2020learning}, especially in biomedical imaging and bioinformatics~\cite{zhao2021mt,lu2022cot,qu2022,pio2022integrating}. The commonly encountered challenges in PPI prediction include label scarcity, where only limited annotations are available for training (semi-supervised learning, SSL), and domain shift, where unseen data (target domain) with different distributions from training data (source domain) is used for testing (unsupervised domain adaptation, UDA). In this regard, model efficiency and generalization would be heavily constrained, limiting the wide real-world applications. Self-ensemble learning~\cite{tepens2017} is one of the most prevalent methods for SSL, which works by enforcing consistency in model predictions from different epochs with the network parameter average~\cite{meanteacher}. Recently, self-ensemble learning has been extended to visual domain adaptation tasks~\cite{choi2019self,zhang2021semi,zhao2022uda}, achieving promising UDA performance. Inspired by these observations, we advance GNN with self-ensemble learning to handle imperfect data for efficient and generalizable PPI prediction.

% \noindent\textbf{Self-ensembling for semi-supervised learning and domain adaptation.}
% Research interest has been drawn to semi-supervised learning (SSL) recently, which learns from both labeled and unlabeled data simultaneously. Self-ensembling is one of the most anticipated direction in SSL~\cite{zhao2020sess}. Self-ensembling methods enhance the model's generalization ability by regularizing ensemble predictions with input perturbations. For instance, temporal ensembling is proposed to form a consensus in model prediction from different epochs~\cite{tepens2017}. Mean teacher improves from the temporal model by substituting the prediction average with the network parameter average~\cite{meanteacher}. SEGCN combines self-ensembling and Graph Convolutional Networks (GCN) for document classification~\cite{luo2020every}. Self-ensembling has also achieved state-of-the-art results in visual domain adaptation tasks~\cite{french2018selfensembling,choi2019self}. In this work, we combine the self-ensembling with the Graph Neural Network (GNN) and adapt it to the PPI prediction task.

% \noindent\textbf{Domain Adaptation}
% Domain shift can serious constraint model performance with heterogeneous training and test data, which is attributed to various reasons,~\emph{e.g.}, different modalities, collection methods, and data formats~\cite{zhang2021semi}. Research efforts have been made to mitigate domain shift in the visual understanding tasks~\cite{zhang2021semi,zhao2022uda}. In our work, we test our method on train-set heterogeneous data to validate its robustness across domains.

\section{Methodology}

\subsection{Task Definition}
% We denote a protein set as $P=\{p_0, p_1, ..., p_n\}$ and its corresponding PPI set as $E=\{e_{ij}={p_i, p_j}|i\neq j, p_i,p_j \in P, I(e_{ij})\in {0, 1}\}$, in which $I$ is the binary PPI indicator function which equals to 1 if the interaction between $p_i$ and $p_j$ is confirmed, or 0 otherwise. The types of PPI are referred to as $C=\{c_0, c_1, ..., c_t\}$ for $t$ PPI types. A PPI $p_{ij}$ may have multiple types $y_{ij} \subseteq C$. To learn the multi-type PPI classification is to find the mapping $F : {p_i, p_j} \rightarrow y_{ij}$. In the semi-supervised setting, the PPI set $E$ is divided into the labeled training set $E_l$, the unlabeled training set $E_u$, and the validation set $E_t$.

Given a set of proteins $P=\{p_0, p_1, ..., p_n\}$ and a set of PPIs $E=\{e_{ij}=\{p_i, p_j\}|i\neq j, p_i,p_j \in P, I(e_{ij})\in {0, 1}\}$, where $I(e_{ij})$ is a binary PPI indicator function that is $1$ if the PPI between proteins $p_i$ and $p_j$ has been confirmed, and $0$ otherwise, the types of PPI can be represented by the label space $C=\{c_0, c_1, ..., c_t\}$ with $t$ different types of interactions, and the labels for a confirmed PPI $e_{ij}$ can be represented as $y_{ij} \subseteq C$. The goal of multi-type PPI learning is to learn a function $f: e_{ij}~\rightarrow \hat{y}_{ij}$ from the training set $E_{train}^s$ such that for any PPI $e_{ij} \in E_{test}^s$, $\hat{y}_{ij}$ is the set of predicted labels for $e_{ij}$. To investigate the efficiency and generalization under complex scenarios beyond the supervised learning setting, we introduce the settings of semi-supervised learning (SSL) and unsupervised domain adaptation (UDA). In the SSL setting, the training datasets consist of limited labeled data $E_{train}^l$ and unlabeled data $E_{train}^u$ due to label scarcity. In the UDA setting, the model trained on $E_{train}^s$ is tested on the unseen data $E_{test}^t$ with different distribution.

\subsection{Overview}
Fig.~\ref{fig:archi} depicts the overview of our proposed SemiGNN-PPI framework. We first construct the multi-graph encoding (MGE) module to effectively leverage available labeled data, which includes a protein graph encoding (PGE) network for exploring protein relations and a label graph encoding (LGE) network for learning label dependencies. To exploit knowledge from unlabeled data, we build a  teacher network with the same architecture as the student network. During teacher-student training, multiple graph consistency constraints at both node and edge levels are utilized to enhance knowledge distillation for self-ensemble multi-graph learning.

\subsection{Multi-Graph Encoding}

\noindent\textbf{Protein-Graph Encoding.}
Early works~\cite{yang2020graph,GNN-PPI} have demonstrated the effectiveness of graph neural networks (GNNs) on PPI prediction. Considering the correlation of PPIs, we use proteins as nodes and PPIs as edges to build the PPI graph $G = (P, E)$. Then, the PPI prediction can be formulated from $f (e_{ij}| p_i,p_j, \theta) \rightarrow \hat{y}_{ij}$ to $f (e_{ij}| G, \theta) \rightarrow \hat{y}_{ij}$. GNNs take the graph structure and sequence-based protein attributes as inputs to model high-level compact representation of the nodes (proteins), denoted by $H~\in \mathbb{R}^{|P|\times d}$ where $h_p=H[p,:]$ is the latent representation of node $p$, and $d$ is the dimensionality of protein features. In general, GNNs follow a recursive neighborhood aggregation scheme to iteratively update the representation of each node by aggregating and transforming the representations of its neighboring nodes. After $l$ iterations, the transformed feature of node $p$ can be denoted as:
\begin{equation}
\label{equ:gnn}
h_p^{(l)} = \phi^{(l)}(h_p^{(l-1)}, f^{(l)}(\{h_p^{(l-1)}: u \in \mathcal{N}_k(p)\})),
\end{equation}
where $\mathcal{N}_k(p)$ denotes the set of $k$-hop neighbors of the node $p$; $f^{(l)}$ and $\phi^{(l)}$ are an aggregation function and a combination function, respectively. Following Graph Isomorphism Network~(GIN)~\cite{xu2018how}, we adopt the summation function to aggregate the representations of neighboring nodes and use the multi-layer perceptrons~(MLPs) to update the aggregated features. Then, the update rule of the hidden node features with a learnable parameter $\epsilon$ in PGE is defined as:
\begin{equation}
\label{equ:gin}
h_p^{(l)}={g}^l((1+\epsilon^l) \cdot h_p^{(l-1)}+\sum\nolimits_{u \in \mathcal{N}_k(p)} h_u^{(l-1)}).
\end{equation}

\noindent\textbf{Label-Graph Encoding.}
In multi-label PPI prediction, correlations exist among different types of interactions,~\emph{i.e.}, some PPI types may appear together frequently while others rarely appear together. Following~\cite{chen2019multi}, we model the interdependencies between different PPI types (labels) using a graph and learn inter-dependent classifiers with Graph Convolutional Network (GCN), which can be directly applied to protein features for multi-type PPI prediction. GCN aims to learn a function $f(\cdot, \cdot)$ on the graph with $t$ nodes. Each GCN layer can be formulated as follows: 

\begin{equation}
h_{c}^{(l+1)}=f({h}_{c}^{(l)} , A), A \in \mathbb{R}^{t \times t},
\end{equation}
where $h_{c}^{(l+1)} \in \mathbb{R}^{t \times d_l^{\prime}}$ and $h_{c}^{(l)} \in \mathbb{R}^{t \times d_l}$ are the learned $d_l^{\prime}$-dimensional node features from current layer and the $d_l$-dimensional node features from previous layer, respectively. $A$ is the corresponding correlation matrix. With the convolutional operation, $f(\cdot, \cdot)$ can be further expressed as: 

\begin{equation}
h_{c}^{(l+1)}=\delta\left(\widehat{A} {h}_{c}^{(l)} W^l\right),
\end{equation}
where $\delta(\cdot)$ is a non-linear function set as LeaklyReLU following~\cite{chen2019multi}, $\widehat{A}$ is the normalized version of $A$ and $W^l \in \mathbb{R}^{{d_l^{\prime} \times d_l}}$ is a transformation matrix. We leverage stacked GCNs to learn inter-dependent classifiers $W$. The first GCN layer takes word embeddings $E_l \in \mathbb{R}^{|t|\times d_l}$ of labels and the correlation matrix $A \in \mathbb{R}^{t \times t}$ as inputs. Considering that PPI type names are semantic, we apply the BioWordVec model~\cite{zhang2019biowordvec} pretrained on the biomedical corpus for generating word embeddings $E_l$ of each PPI type to better capture their semantics. To construct the label correlation matrix $A$, we compute the conditional probability of different labels within the training dataset. To avoid noises and over-smoothing, we binarize $A$ with a threshold $\tau$ and then re-weight it with a weight $p$ to obtain $\widehat{A}$.

\noindent\textbf{Multi-Graph Based Classifier Learning.}
By applying the learned classifiers $W = \left\{w_i \right\}_{i=1}^{t}$ from label graph encoding (LGE) to the learned representations from protein graph encoding (PGE) for the PPI $e_{ij}$, we can obtain the predicted scores $\hat{y}_{ij}$, expressed as:

\begin{equation}
\hat{y}_{ij}= W(h_{p_i} \cdot h_{p_j}).
\end{equation}
We use the traditional multi-label classification loss function to update the whole network in an end-to-end manner. The loss function can be written as:

\begin{equation*}
\mathcal{L}_{sup}=\sum_{c=1}^t\left(y^c \log \left(\sigma\left(\hat{y}^c\right)\right)+\left(1-y^c\right) \log \left(1-\sigma\left(\hat{y}^c\right)\right)\right),
\end{equation*}
where $\sigma(\cdot)$ is the sigmoid function. Our model learns the aggregated features by combining protein neighbors and models the label correlations by learning inter-dependent classifiers simultaneously to improve the model generalization. In multi-graph learning, the learned classifiers are expected to be neighborhood aware at both feature and label levels.

\subsection{Self-ensemble Graph Learning}
To leverage unlabeled data, we adopt the mean teaching architecture for unsupervised learning, as shown in Fig.~\ref{fig:archi}.
We construct a teacher network $f_t$ with the same architecture as the student network $f_s$ based on self-ensembling~\cite{meanteacher}. Specifically, in each training iteration $\text{k}$, we update the teacher model weights $\theta^{\prime}$ with the exponential moving average (EMA)  weights of the student model $\theta$ by leveraging the momentum updating mechanism:
\begin{equation}
\theta_{\text{k}}^{\prime}=m \theta_{\text{k}-1}^{\prime}+(1-m) \theta_{\text{k}},
\end{equation}
where $m$ is momentum. During training, the student model is encouraged to be consistent with the teacher predictions for the inputs with different augmentations. Because of the non-euclidean graph structure, image augmentations such as crop and rotation cannot be directly applied to graphs. To facilitate self-ensemble graph learning, we construct two graph data augmentation methods at both the edge and node levels,~\emph{i.e.}, \textbf{Edge Manipulation} and \textbf{Node Manipulation} to augment graph topological and attribute information~\cite{you2020graph}. 
\textbf{Edge Manipulation (EM):} To improve the robustness against connectivity variations, we randomly replace a certain percentage of edges in the input to the student and teacher models, since some edges (PPIs) between different nodes (proteins) may be unidentified or wrong in experimental procedures. Specifically, we follow an i.i.d. uniform distribution to randomly replace $em_s\%$ and $em_t\%$ of edges in the input to the student and the teacher, respectively. Different from~\cite{you2020graph}, we replace the dropped edge by linking the node with one of its neighbor's neighboring nodes for maintaining global structural information,~\emph{i.e.,} node $p_s$ with a dropped edge connecting to node $p_t$ could be linked to $p_u\in\{p_u|e_{ut}=1\}$. 
\textbf{Node Manipulation (EM):} To improve the robustness against attribute missing, we randomly remove $nm_s\%$ and $nm_t\%$ of node features, mask them with zeros and feed them into the student and teacher models respectively, to expect the model to effectively learn the features even in the presence of missing attribute information.
We construct two graph views with augmentations above to feed the student and teacher networks separately, and encourage them to generate consistent predictions using $\ell_2$ loss:

\begin{equation}
\mathcal{L}_{con}= \|f_t (E_u| G, \theta_{k}^{\prime}, \xi^{\prime})  - f_s (E_u| G, \theta_{k}, \xi) \|_2,
\end{equation}
where $E_u$ is unlabeled PPIs in a batch. $\xi^{\prime}$ and $\xi$ are different augmentation operations. We randomly comprise the different augmentations in our experiments to avoid overfitting and improve model generalization.

\begin{table*}[ht]
\centering
\small
\setlength\tabcolsep{5pt}
\scalebox{0.8}{
\begin{tabular}{cl|ccc|ccc|ccc} 
\hline
\multicolumn{2}{c|}{\multirow{2}{*}{Method}} & \multicolumn{3}{c|}{SHS27k}                                                    & \multicolumn{3}{c|}{SHS148k}                                                & \multicolumn{3}{c}{STRING}                                        \\ 
\cline{3-11}
\multicolumn{2}{c|}{}                        & \multicolumn{1}{l}{Random} & DFS                     & BFS                     & Random                  & DFS                     & BFS                     & Random       & DFS                     & BFS                      \\ 
\hline
\multirow{2}{*}{ML}    & RF                  & $78.45_{0.88}$               & $35.55_{2.22}$            & $37.67_{1.57}$            & $82.10_{0.20}$                & $43.26_{3.43}$            & $38.96_{1.94}$            & $88.91_{0.08}$                & $70.80_{0.45}$            & $55.31_{1.02}$             \\
                       & LR                  & $71.55_{0.93}$               & $48.51_{1.87}$            & $43.06_{5.05}$            & $67.00_{0.07}$                & $51.09_{2.09}$            & $47.45_{1.42}$            & $67.74_{0.16}$                & $61.28_{0.53}$            & $50.54_{2.00}$             \\ 
\hline
\multirow{3}{*}{DL}    & DPPI                & $73.99_{5.04}$               & $46.12_{3.02}$            & $41.43_{0.56}$            & $77.48_{1.39}$                & $52.03_{1.18}$            & $52.12_{8.70}$            & $94.85_{0.13}$                & $66.82_{0.29}$            & $56.68_{1.04}$             \\
                       & DNN-PPI             & $77.89_{4.97}$               & $54.34_{1.30}$            & $48.90_{7.24}$            & $88.49_{0.48}$                & $58.42_{2.05}$            & $57.40_{9.10}$            & $83.08_{0.11}$                & $64.94_{0.93}$            & $53.05_{0.82}$             \\
                       & PIPR                & $83.31_{0.75}$               & $57.80_{3.24}$            & $44.48_{4.44}$            & $90.05_{2.59}$                & $63.98_{0.76}$            & $61.83_{10.23}$           & $94.43_{0.10}$                & $67.45_{0.34}$            & $55.65_{1.60}$             \\ 
\hline
\multirow{2}{*}{Graph} & GNN-PPI             & $87.91_{0.39}$               & $74.72_{5.26}$            & $63.81_{1.79}$            & $92.26_{0.10}$                & $82.67_{0.85}$            & $71.37_{5.33}$            & $95.43_{0.10}$                & $91.07_{0.58}$            & $78.37_{5.40}$             \\
                       & GNN-PPI*            & $88.87_{0.23}$               & $75.68_{3.95}$            & $68.84_{3.16}$            & $92.13_{0.10}$                & $83.77_{1.34}$            & $69.02_{3.07}$            & $94.94_{0.17}$               & $90.62_{0.23}$            & $79.76_{2.43}$             \\ 
\hline
M-Graph                & SemiGNN-PPI         & $\mathbf{89.51_{0.46}}$    & $\mathbf{78.32_{3.15}}$ & $\mathbf{72.15_{2.87}}$ & $\mathbf{92.40_{0.22}}$     & $\mathbf{85.45_{1.17}}$ & $\mathbf{71.78_{3.56}}$ & $\mathbf{95.57_{0.08}}$                & $\mathbf{91.23_{0.26}}$ & $\mathbf{80.84_{2.05}}$  \\
\hline
\end{tabular}
}
\caption{Performance of SemiGNN-PPI and baseline methods over different datasets and data partition schemes. GNN-PPI: reported results in the original paper. GNN-PPI$^*$: reproduced GNN-PPI results. The scores are presented in the format of $\mathrm{mean_{std}}$.}
\label{tab:benchmark}
\end{table*}

\subsection{Graph Consistency Constraint}
The consistency regularization enforces instance-wise invariance on the prediction space towards different augmentations on the same input, describing the PPI interactions between samples. 
For the graph-based PPI prediction task, we also need to optimize the model in the feature space, as protein nodes in the testing set differ from the training set and PPI is performed as the relationships between proteins by feature representations extracted from neighboring proteins. Therefore, we model the fine-grained structural protein-protein relations in the feature embedding space~\cite{ma2022distilling}. We denote the features extracted from protein-graph encoding as $z_{s}$ and $z_{t}$ for the student and teacher networks, respectively. \textbf{Edge matching:} We construct the student embedding graph $G_e^s$ and the teacher embedding graph $G_e^t$ by calculating all pairwise Pearson's correlation coefficient (PCC) between nodes in the same batch. Then, we enforce the student network to encode consistent instance-wise correlations with the teacher network in the embedding feature space by applying the edge matching loss: 

\begin{equation}
\mathcal{L}_{edge}= ||\text{Adj}(G_e^s) - \text{Adj}(G_e^t)||_2,
\label{eq:edge_matching_loss}
\end{equation}
where $\text{Adj}$ refers to the adjacency matrix. \textbf{Node matching:} We further formulate the edge embedding graph $G_e^{st}$ by calculating all pairwise PCC between student encoding $z_{s}$ and teacher encoding $z_{t}$ in the same batch. To explicitly align encoding of the same protein from the teacher and the student network, we design a node matching loss:
\begin{equation}
\mathcal{L}_{node}= ||\text{diag}(\text{Adj}(G_e^{st})) - \text{diag}(I)||_2,
\label{eq:node_matching_loss}
\end{equation}
where $\text{diag}$ is an operator to create a block-diagonal matrix with the off-diagonal elements of $0$, and $I$ refers to the identity matrix. In this regard, we jointly leverage labeled and unlabeled data with graph learning in both protein and label spaces and consistency regularization in both prediction and feature spaces for PPI prediction.
The overall objective function is defined as:
\begin{equation}
\mathcal{L}= \mathcal{L}_{sup} + \lambda_{con}\mathcal{L}_{con} + \lambda_{edge}\mathcal{L}_{edge} + \lambda_{node}\mathcal{L}_{node},
\label{eq:overall_loss}
\end{equation}
 where $\lambda_{con}$, $\lambda_{edge}$ and $\lambda_{node}$ are scaling factors for $\mathcal{L}_{con}$, $\mathcal{L}_{edge}$ and $\mathcal{L}_{node}$, respectively.
\begin{table*}[ht]
\centering
\small
\setlength\tabcolsep{2.5pt}
\scalebox{0.76}{
\begin{tabular}{l|cccc|cccc|cccc} 
\toprule
\multirow{2}{*}{Method} & \multicolumn{4}{c|}{STRING}                           & \multicolumn{4}{c|}{SHS148k}                           & \multicolumn{4}{c}{SHS27k}                              \\
                        & 5\%         & 10\%        & 20\%        & 100\%       & 5\%          & 10\%        & 20\%        & 100\%       & 5\%         & 10\%         & 20\%        & 100\%        \\ 
\hline\hline
\multicolumn{13}{c}{Partition Scheme = Random}                                                                                                                                  \\ 
\hline
GNN-PPI  & $89.94_{0.29}$ & $92.38_{0.51}$ & $93.30_{0.56}$ & $94.94_{0.17}$ & $79.19_{0.67}$  & $82.86_{0.49}$ & $86.67_{0.22}$ & $92.13_{0.10}$ & $52.04_{3.32}$ & $60.28_{12.26}$ & $79.44_{1.19}$ & $88.87_{0.23}$  \\ 
\hline
Ours     & $\mathbf{90.55_{0.10}}$ & $\mathbf{92.66_{0.59}}$ & $\mathbf{93.90_{0.41}}$ & $\mathbf{95.57_{0.08}}$ & $\mathbf{79.50_{0.31}}$  & $\mathbf{83.48_{0.30}}$ & $\mathbf{87.38_{0.24}}$ & $\mathbf{92.40_{0.22}}$ & $\mathbf{57.97_{1.13}}$ & $\mathbf{62.67_{11.26}}$ & $\mathbf{81.01_{0.47}}$ & $\mathbf{89.51_{0.46}}$ \\ 
\hline\hline
\multicolumn{13}{c}{Partition Scheme = DFS}      \\ 
\hline
GNN-PPI                 & $86.60_{0.37}$ & $87.91_{0.30}$ & $89.42_{0.46}$ & $90.62_{0.23}$ & $68.77_{11.20}$ & $78.36_{2.23}$ & $80.96_{1.61}$ & $83.77_{1.34}$ & $53.41_{1.64}$ & $58.43_{2.27}$  & $65.73_{4.18}$ & $75.68_{3.95}$  \\ 
\hline
Ours                    & $\mathbf{87.54_{0.06}}$ & $\mathbf{88.98_{0.26}}$ & $\mathbf{90.23_{0.12}}$ & $\mathbf{91.23_{0.26}}$ & $\mathbf{69.94_{9.57}}$  & $\mathbf{81.12_{0.98}}$ & $\mathbf{83.63_{0.86}}$ & $\mathbf{85.45_{1.17}}$ & $\mathbf{58.48_{1.11}}$ & $\mathbf{61.18_{1.98}}$  & $\mathbf{70.31_{2.38}}$ & $\mathbf{78.32_{3.15}}$ \\ 
\hline\hline
\multicolumn{13}{c}{Partition Scheme = BFS}                                               \\ 
\hline

GNN-PPI                 & $71.35_{4.67}$ & $74.94_{2.35}$ & $79.99_{2.75}$ & $79.76_{2.43}$ & $61.42_{3.29}$  & $62.51_{3.07}$ & $67.10_{3.48}$ & $69.02_{3.07}$ & $57.93_{4.11}$ & $56.84_{12.19}$ & $61.18_{6.58}$ & $68.84_{3.16}$  \\ 
\hline
Ours                    & $\mathbf{73.35_{4.90}}$ & $\mathbf{76.94_{2.53}}$ & $\mathbf{81.39_{2.44}}$ & $\mathbf{80.84_{2.05}}$ & $\mathbf{64.86_{2.97}}$  & $\mathbf{68.76_{1.62}}$ & $\mathbf{71.06_{3.35}}$ & $\mathbf{71.78_{3.56}}$ & $\mathbf{60.15_{2.09}}$ & $\mathbf{66.13_{2.01}}$  & $\mathbf{67.69_{8.47}}$ & $\mathbf{72.15_{2.87}}$  \\
\bottomrule
\end{tabular}}
\caption{Performance comparison of different methods under different label ratios. The scores are presented in the format of $\mathrm{mean_{std}}$. 
}
\label{tab:label_efficiency}
\end{table*}

\begin{table*}[ht]
\centering
\small
\setlength\tabcolsep{6pt}
\scalebox{0.75}{
\begin{tabular}{c|c|ccc|cc|cc} 
\toprule
Method      & \% Labels            & \multicolumn{3}{c|}{Random Partition}                     & \multicolumn{2}{c|}{DFS Partition}    & \multicolumn{2}{c}{BFS Partition}  \\ 
\hline
            & \multirow{3}{*}{100} & BS (92.66\%)        & ES (6.95\%)         & NS(0.39\%)          & ES (75.95\%)        & NS(24.05\%)         & ES (85.70\%) & NS(14.30\%)             \\ 
\cmidrule(lr){3-3}\cmidrule(lr){4-4}\cmidrule(lr){5-5}\cmidrule(lr){6-6}\cmidrule(lr){7-7}\cmidrule(lr){8-8}\cmidrule(lr){9-9}
GNN-PPI     &                      & 89.17            & 72.44            & 50.00             & 77.81     & 63.44       & 71.03            & 44.80          \\
SemiGNN-PPI &                      & $\mathbf{89.68}$ & $\mathbf{72.93}$  & $50.00$        & $\mathbf{81.75}$   & $\mathbf{66.32}$  & $\mathbf{75.14}$          & $\mathbf{57.00}$                \\ 
\hline
            & \multirow{3}{*}{20}  & BS (73.18\%)        & ES (24.98\%)        & NS (1.84\%)         & ES (72.87\%)        & NS (27.13\%)        & ES (47.71\%) & NS (52.29\%)            \\ 
\cmidrule(lr){3-3}\cmidrule(lr){4-4}\cmidrule(lr){5-5}\cmidrule(lr){6-6}\cmidrule(lr){7-7}\cmidrule(lr){8-8}\cmidrule(lr){9-9}
GNN-PPI     &                      & 83.46            & 70.10            & 43.68            & 64.40            & 54.21            & $\mathbf{59.04}$     & 66.33                \\
SemiGNN-PPI &                      & $\mathbf{84.09}$ & $\mathbf{71.95}$           & $\mathbf{45.78}$          & $\mathbf{73.30}$          & $\mathbf{55.46}$           & $58.10$     & $\textbf{73.82}$               \\ 
\hline
            & \multirow{3}{*}{10}  & BS (55.80\%)        & ES (38.03\%)        & NS (6.16\%)         & ES (63.36\%)        & NS (36.64\%)        & ES (41.14\%) & NS (58.86\%)            \\ 
\cmidrule(lr){3-3}\cmidrule(lr){4-4}\cmidrule(lr){5-5}\cmidrule(lr){6-6}\cmidrule(lr){7-7}\cmidrule(lr){8-8}\cmidrule(lr){9-9}
GNN-PPI     &                      & 79.64             & 69.64            & 38.41            & 56.13            & 53.85            & 36.02     & 47.89                 \\
SemiGNN-PPI &                      & $\mathbf{80.22}$ & $\mathbf{70.33}$ & $\mathbf{41.67}$ & $\mathbf{61.07}$  & $\mathbf{57.90}$ & $\mathbf{57.39}$    & $\mathbf{72.73}$              \\ 
\hline
            & \multirow{3}{*}{5}   & BS (38.16\%)        & ES (47.61\%)        & NS (14.23\%)        & ES (46.63\%)        & NS (53.37\%)        & ES (43.18\%) & NS (56.82\%)            \\ 
\cmidrule(lr){3-3}\cmidrule(lr){4-4}\cmidrule(lr){5-5}\cmidrule(lr){6-6}\cmidrule(lr){7-7}\cmidrule(lr){8-8}\cmidrule(lr){9-9}
GNN-PPI     &                      & 53.43             & 44.33            & 40.64            & 53.85             & 49.62            & 56.10     & 51.95                \\
SemiGNN-PPI &         & $\mathbf{59.76}$ & $\mathbf{57.82}$ & $\mathbf{42.71}$ & $\mathbf{58.25}$ & $\mathbf{56.25}$          & $\mathbf{58.18}$   & $\mathbf{58.60}$              \\
\bottomrule
\end{tabular}}
\caption{Analysis on performance between GNN-PPI and SemiGNN-PPI over BS, ES, and NS subsets in the SHS27k dataset. The ratios of the subsets are annotated in brackets. The BS subsets are empty under DFS and BFS partitions and are omitted for brevity.}
\label{tab:in_depth}
\end{table*}

\section{Experiment}

\subsection{Dataset}
We perform extensive experiments on three datasets,~\emph{i.e.,} STRING, SHS148k, and SHS27k. First, we use the multi-label PPI data of Homo sapiens from the STRING database~\cite{szklarczyk2019string} for training and evaluation, including $15,355$ proteins and $593,397$ PPIs. The PPIs are annotated with $7$ types,~\emph{i.e.,} Activation, Binding, Catalysis, Expression, Inhibition, Post-translational modification (Ptmod), and Reaction. Each PPI is labeled with at least one of them. Moreover, we use two subsets of Homo sapiens PPIs from STRING,~\emph{i.e.}, SHS27k, and SHS148k~\cite{PIPR}, to further validate the proposed approach. SHS27k contains $1,690$ proteins and $7,624$ PPIs, while SHS148k contains $5,189$ proteins and $44,488$ PPIs.

\subsection{Experimental Details}
% \subsubsection{Experimental Settings and Metrics}
% Pls ref that IJCAI paper and input the three evaluation frameworks, i.e., BFS, DFS, Random. Also include the label scarcity scenario, and domain shift scenario. may ref some SSL / DA papers. pls include your refs as comments.

\noindent\textbf{Experimental Settings.} We follow partition algorithms in GNN-PPI~\cite{GNN-PPI}, including random, breath-first search (BFS), and depth-first search (DFS) to split the trainsets and testsets. For in-depth analysis, PPIs in the testset can be divided into~\textbf{BS subset} (both proteins of the PPI are present in the labeled trainset),~\textbf{ES subset} (either one protein of the PPI is present in the labeled trainset), and~\textbf{NS subset} (neither of the proteins is present in the labeled trainset). The BFS and DFS partition schemes create more challenging paradigms than the random partitioning by including more ES and NS proteins in the testsets for the inter-novel protein interactions~\cite{GNN-PPI}. In fully supervised experiments, we select $20\%$ of the whole dataset for testing using the partition schemes mentioned above and use the rest for training. To simulate the label scarcity scenario, we randomly select  $5\%$, $10\%$, and $20\%$ samples from the trainset as the labeled data while keeping the rest as the unlabeled data. To assess the generalization capacity of our method, we evaluate our method trained with one dataset on another dataset,~\emph{i.e.}, a trainset-heterologous testset.

% In a random-partitioned trainset, we randomly select labeled PPIs and keep the rest unlabeled. In a BFS-partitioned trainset, we keep the first $n$ PPIs found during BFS as labeled and the rest as unlabeled, where $n$ depends on the different proportions of labeled data in the experiments. The same approach applies to a DFS-partitioned trainset. 
\noindent\textbf{Evaluation Metrics.} We use the F1 score to evaluate the model performance for multi-label PPI prediction. The score is micro-averaged over all $7$ classes. The means and variances of F1 scores over three repeated experiments are reported as results, formatted as $\mathrm{mean_{std}}$.

\noindent\textbf{Model Training.} 
1)~\textit{Base train}: We follow GNN-PPI~\cite{GNN-PPI} for protein-independent encoding to extract protein features from protein sequences as inputs to our framework. We initialize the multi-graph encoding network using the labeled data for $300$ epochs with an initial learning rate of $0.001$ and the Adam optimizer.
2)~\textit{Joint train}: Then, we train the self-ensemble graph learning framework on both labeled and unlabeled trainsets for $300$ epochs. For label graph construction, we select the binarization threshold $\tau=0.05$ and the re-weighting factor $p=0.25$. We randomly comprise the different manipulations in our experiments to avoid overfitting and improve model generalization during joint training. For manipulation ratios, we use higher
ratios for the student inputs so that the student can better distill knowledge from the teacher during self-ensemble learning. More specifically, the edge manipulation ratios $em_s\%$ and $em_t\%$ are fixed at $10\%$ and $5\%$, respectively. The node manipulation rates $nm_s\%$, and $nm_t\%$ are set to $10\%$ and $5\%$, respectively. To scale the components of the loss function, we set the value of $\lambda_{con}$, $\lambda_{edge}$ and $\lambda_{node}$ as $0.02$, $0.01$ and $0.003$, respectively. More details are shown in Supplementary Material.

\noindent\textbf{Baseline Methods.}
We compare SemiGNN-PPI with several representative methods in PPI prediction, including: \textbf{Machine Learning (ML)} methods include RF~\cite{RF-PPI} and LR~\cite{LR-PPI}, which take commonly handcrafted protein features including AC~\cite{SVM-PPI} and CTD~\cite{DeepPPI} as inputs. \textbf{Deep Learning (DL)} approaches include DNN-PPI~\cite{DNN-PPI}, PIPR~\cite{PIPR}, and GNN-PPI~\cite{GNN-PPI}, which take amino acid sequence-based features as inputs (More details are illustrated in Appendix). It is noted that GNN-PPI adopts graph learning to leverage protein correlations, achieving state-of-the-art performance on multi-type PPI prediction. In this regard, we extensively compare our method with GNN-PPI in different scenarios and settings.
% \begin{itemize}
%     \item \textbf{Machine Learning (ML)} methods include RF~\cite{RF-PPI} and LR~\cite{LR-PPI}, which take commonly handcrafted protein features including AC~\cite{SVM-PPI} and CTD~\cite{DeepPPI} as inputs.
%     \item \textbf{Deep Learning (DL)} approaches include DNN-PPI~\cite{DNN-PPI} and PIPR~\cite{PIPR} and GNN-PPI~\cite{GNN-PPI}, which take amino acid sequence-based features as inputs (More details are illustrated in Appendix). It is noted that GNN-PPI adopts graph learning to leverage protein correlations, achieving state-of-the-art performance on multi-type PPI prediction. In this regard, we extensively compare our method with GNN-PPI in different scenarios and settings.
% \end{itemize}

\subsection{Results and Analysis}
\noindent\textbf{Benchmark Analysis.}
In Table~\ref{tab:benchmark}, we compare our methods with other baseline methods under different partition schemes and various datasets. It is observed that graph-based methods,~\emph{i.e.}, GNN-PPI and SemiGNN-PPI outperform other ML and DL methods, even under more challenging BFS and DFS partitions with more unseen proteins. It can be attributed to graph learning, which can better capture correlations between proteins despite the existence of more unknown proteins. Furthermore, our method incorporates multiple graphs~(M-Graph) for feature learning, achieving state-of-the-art performance in multi-type PPI prediction. Especially, under challenging evaluations with small datasets,~\emph{e.g.}, SHS27k-DFS, our method achieves much higher F1 scores than GNN-PPI, since self-ensemble graph learning can effectively improve the model robustness against complex scenarios. Moreover, the number
of parameters is 1.09M (GNN-PPI) and 1.13M (ours), and the inference time on SHS27k is 0.050s (GNN-PPI) and 0.058s (ours). GNN-PPI and our method have comparable performance in the two metrics, showing the scalability of the proposed method.

% To evaluate the overall performance of our approach, we present the benchmark results in Table~\ref{tab:benchmark}. It is clear that our method achieves higher F1 scores in most conditions, indicating the consistency and robustness of our approach in different datasets and partition schemes. Moreover, the proposed method outperforms the baselines despite the fact that the benchmark is conducted in a fully supervised manner. We attribute the performance increment to 1) the label correlations captured by the label graph encoding and 2) the regularization effect of self-ensemble that further optimizes the decision boundary from supervised learning.

\begin{table*}[!ht]
\centering
\small
\setlength\tabcolsep{11pt}
\scalebox{0.8}{
\begin{tabular}{cccccccc} 
\toprule
\multirow{2}{*}{PPI Type} & \multirow{2}{*}{Type Ratio}  & \multicolumn{2}{c}{Random Partition} & \multicolumn{2}{c}{DFS Partition}  & \multicolumn{2}{c}{BFS Partition}   \\ 
\cline{3-8}
                          &                         & GNN-PPI    & SemiGNN-PPI   & GNN-PPI    & SemiGNN-PPI & GNN-PPI    & SemiGNN-PPI  \\ 
\hline
Reaction                  & $40.61$\% & $89.58_{0.15}$ & $\mathbf{90.16_{0.43}}$    & $81.90_{1.65}$ & $\mathbf{85.86_{0.71}}$  & $61.62_{1.29}$ & $\mathbf{64.92_{5.73}}$   \\
Binding                   & $52.71\%$ & $88.28_{0.48}$ & $\mathbf{89.46_{0.57}}$    & $83.52_{1.41}$ & $\mathbf{86.39_{0.67}}$  & $70.00_{4.10}$ & $\mathbf{72.43_{6.33}}$   \\
Ptmod                     & $20.99\%$  & $87.04_{0.29}$ & $\mathbf{87.42_{0.33}}$    & $77.94_{1.67}$ & $\mathbf{82.99_{1.44}}$  & $65.92_{5.52}$ & $\mathbf{71.32_{5.04}}$   \\
Activation                & $42.51\%$ & $85.15_{0.38}$ & $\mathbf{85.26_{0.46}}$    & $73.48_{2.74}$ & $\mathbf{77.95_{1.19}}$  & $67.44_{8.43}$ & $\mathbf{68.04_{8.06}}$   \\
Inhibition                & $20.20\%$  & $87.21_{0.18}$ & $\mathbf{88.09_{0.31}}$    & $72.46_{1.11}$ & $\mathbf{78.12_{2.62}}$  & $60.20_{4.62}$ & $\mathbf{67.71_{7.21}}$   \\
Catalysis                 & $44.67\%$ & $89.36_{0.44}$ & $\mathbf{90.35_{0.31}}$    & $82.30_{0.80}$ & $\mathbf{85.77_{1.29}}$  & $65.70_{4.42}$ & $\mathbf{73.39_{6.33}}$   \\
Expression                & $7.69\%$  & $\mathbf{47.85_{0.79}}$ & $46.99_{0.22}$    & $\mathbf{34.96_{3.74}}$  & $32.45_{5.96}$  & $\mathbf{31.81_{6.87}}$  & $28.99_{4.90}$   \\ 
\hline
Macro-Average   & -   & $82.07_{0.39}$ & $\mathbf{82.53_{0.38}}$    & $72.37_{1.87}$ & $\mathbf{74.16_{2.09}}$  & $60.38_{5.03}$ & $\mathbf{63.29_{5.29}}$   \\
Micro-Average   & -   & $86.67_{0.22}$ & $\mathbf{87.38_{0.24}}$    & $80.96_{1.61}$ & $\mathbf{83.63_{0.86}}$  & $67.10_{3.48}$ & $\mathbf{71.06_{3.35}}$   \\

\bottomrule
\end{tabular}
}
\caption{Per-class results in the SHS148k dataset with $20\%$ training labels. The type ratios are calculated over the whole dataset.}
\label{tab:percls}
\end{table*}

\noindent\textbf{Label Efficiency.}
To demonstrate the feasibility of our method under the label scarcity scenario, we present experimental results under different label ratios in Table~\ref{tab:label_efficiency}. We can see that GNN-PPI receives severe performance degradation with fewer labels. In comparison, our method achieves better performance under all scenarios with different datasets, label ratios, and partition schemes. Remarkably, our method under some scenarios,~\emph{e.g.}, SHS148k-BFS-$20\%$ can achieve comparable performance with GNN-PPI using $100\%$ labeled data, indicating the annotation efficiency of our method. To further analyze the model performance on inter-novel-protein interaction prediction, we make an in-depth performance comparison between GNN-PPI and SemiGNN-PPI in the different subsets (BS/ES/NS) of the testset. As shown in Table~\ref{tab:in_depth}, the BS subset comprises most of the whole testset under the random partition, which cannot reflect the prediction performance on the inter-novel-protein interactions. In contrast, the proportions of ES and NS subsets increase under label scarcity and other partition schemes; in these settings, SemiGNN-PPI consistently outperforms GNN-PPI in both ES and NS subsets by a large margin, which demonstrates the effectiveness of SemiGNN-PPI for inter-novel-protein interaction prediction.

% In Table~\ref{tab:label_efficiency}, we demonstrate experimental results with different ratios of labeled training data to study the model performance under label scarcity., SemiGNN-PPI outperforms baselines in all combinations of the dataset, percentage of labels, and the partition scheme, especially in the challenging BFS and DFS splits with more unseen proteins in the testset~\cite{GNN-PPI}. Moreover, it is noticeable that in some settings,~\emph{e.g.}, SHS148k-BFS-$20\%$, our approach can achieve results on par with GNN-PPI~\cite{GNN-PPI} using $100\%$ of labeled data, indicating that our model effectively learns from the unlabeled data. 

% In Table.~\ref{tab:in_depth}, we analyze the model performance from the perspective of BS/ES/NS subsets in the testset. It can be observed that the limited amount of labels and the DFS/BFS partitions jointly create challenging scenarios by increasing the ratio of ES/NS PPIs in the testset. Despite the large proportion of PPI with novel proteins, our approach consistently outperforms the baseline in ES/NS PPIs by a large margin. In summary, the experimental results confirm the consistently superior performance of the proposed SemiGNN-PPI in label-scarce scenarios.

\noindent\textbf{Performance on Different PPI Types.}
To study the per-class prediction performance, we present the model performance on different PPI types with corresponding type ratios in Table~\ref{tab:percls}. It is observed that the PPI types are unbalanced with some under-represented types, such as Ptmod, Inhibition, and Expression. Nevertheless, SemiGNN-PPI outperforms GNN-PPI on most PPI types, especially for relatively imbalanced types (82.99 vs. 77.94 in Ptmod-DFS and 67.71 vs. 60.20 in Inhibition-BFS). It is noted that lower performance is achieved with our method in the type Expression, which could be due to inaccurate label correlations captured with extremely low co-occurrence with other labels, which is still a direction to explore in our future work. 

\begin{figure}[ht]
    \centering
    \includegraphics[width = 0.47\textwidth]{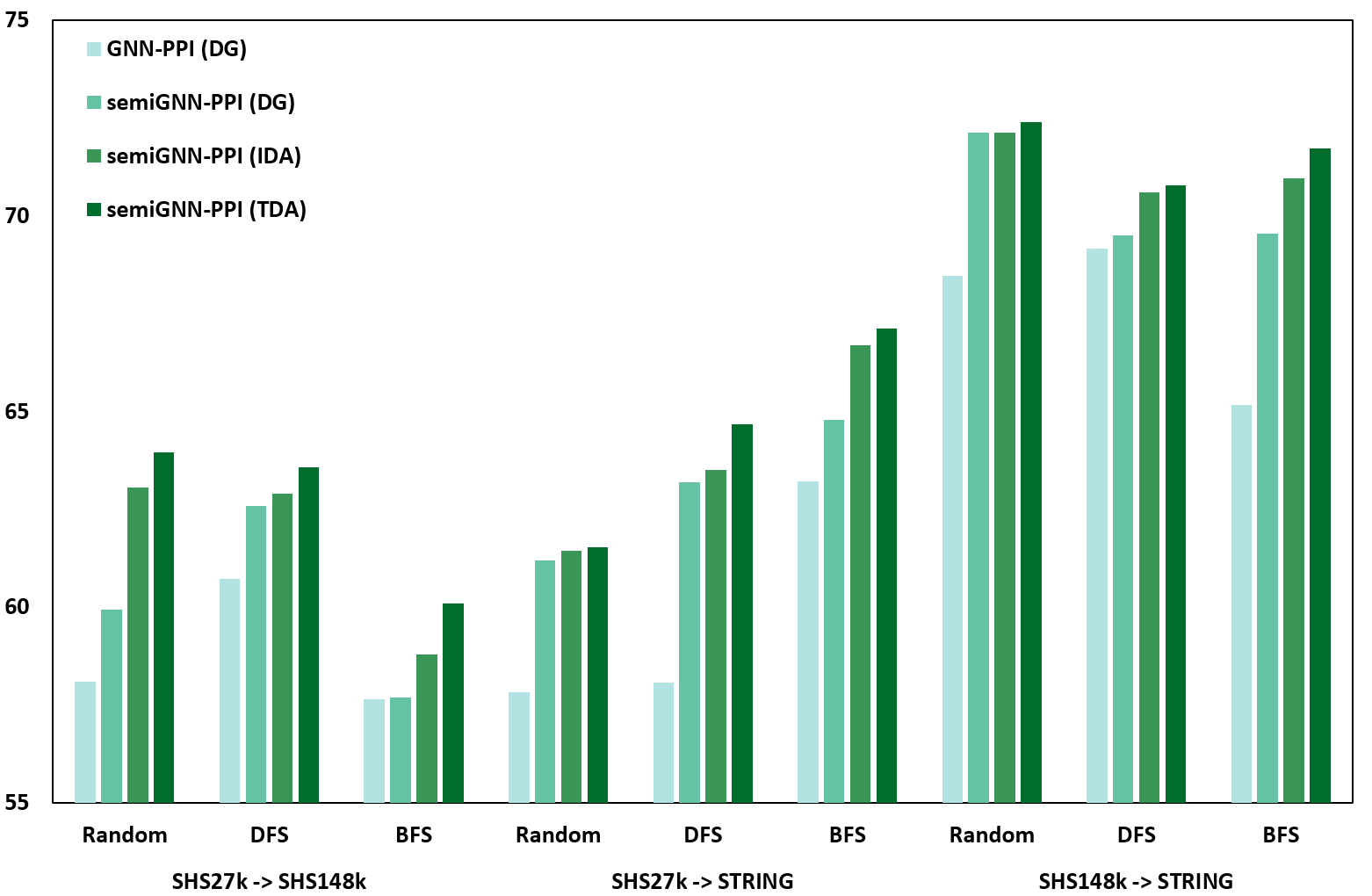}
    \caption{Performance comparison on trainset-heterologous testsets. DG: domain generalization. IDA: inductive domain adaptation. TDA: transductive domain adaptation.}
    \label{fig:model_gen}
\end{figure}

\noindent\textbf{Model Generalization.}
To access the generalization capability of the proposed method, we test the model trained using small datasets,~\emph{e.g.,} SHS27k on big datasets,~\emph{e.g.,} STRING in three evaluation settings: 1)~\textit{Domain Generalization (DG)}: The model is directly tested on the unseen dataset. 2)~\textit{Inductive Domain Adaptation (IDA)}: The model has access to unlabeled training data in the trainset-heterologous dataset during training. 3)~\textit{Transductive domain adaptation}: The model has access to the whole unlabeled trainset-heterologous dataset during training. In Fig.~\ref{fig:model_gen}, we can observe that our method outperforms GNN-PPI in all partition schemes when tested on unseen datasets. Moreover, our model can effectively leverage unlabeled data, achieving better adaptation performance in both inductive and transductive setups.

% To explore the generalization capability of the proposed method, we present experimental results on trainset-heterologous testset in Table.~\ref{tab:model_gen}. Specifically, we probe $3$ scenarios: 1)~\textit{Model generalization}: The model is trained only on the source trainset. 2)~\textit{Inductive domain adaptation}: The model is trained on the source trainset and fine-tuned on the unlabeled target trainset. 3)~\textit{Transductive domain adaptation}: The model is trained on the source trainset and fine-tuned on the entire unlabeled target dataset. In Table.~\ref{tab:model_gen}, we can observe that SemiGNN-PPI consistently outperforms GNN-PPI~\cite{GNN-PPI} in all settings. Moreover, our approach is further enhanced with inductive and transductive fine-tuning. The results confirm our assumption that compared with the baseline GNN-PPI~\cite{GNN-PPI}, our model can effectively leverage the unlabeled data from the target domain for optimized performance. 

% To inspect the classification performance of our approach, we show per-class results in Table~\ref{tab:percls}. It is obvious that the proportions of PPI types are unbalanced, with $3$ under-represented types of PPIs (Ptmod, Inhibition, and Expression). Despite the class imbalance, SemiGNN-PPI improves from the baseline in most PPI types except the Expression Type, which could be due to the inaccurate label correlations captured by the model based on its low co-occurrence with other labels. 

\begin{figure}[ht]
    \centering
    \includegraphics[width = 0.4\textwidth]{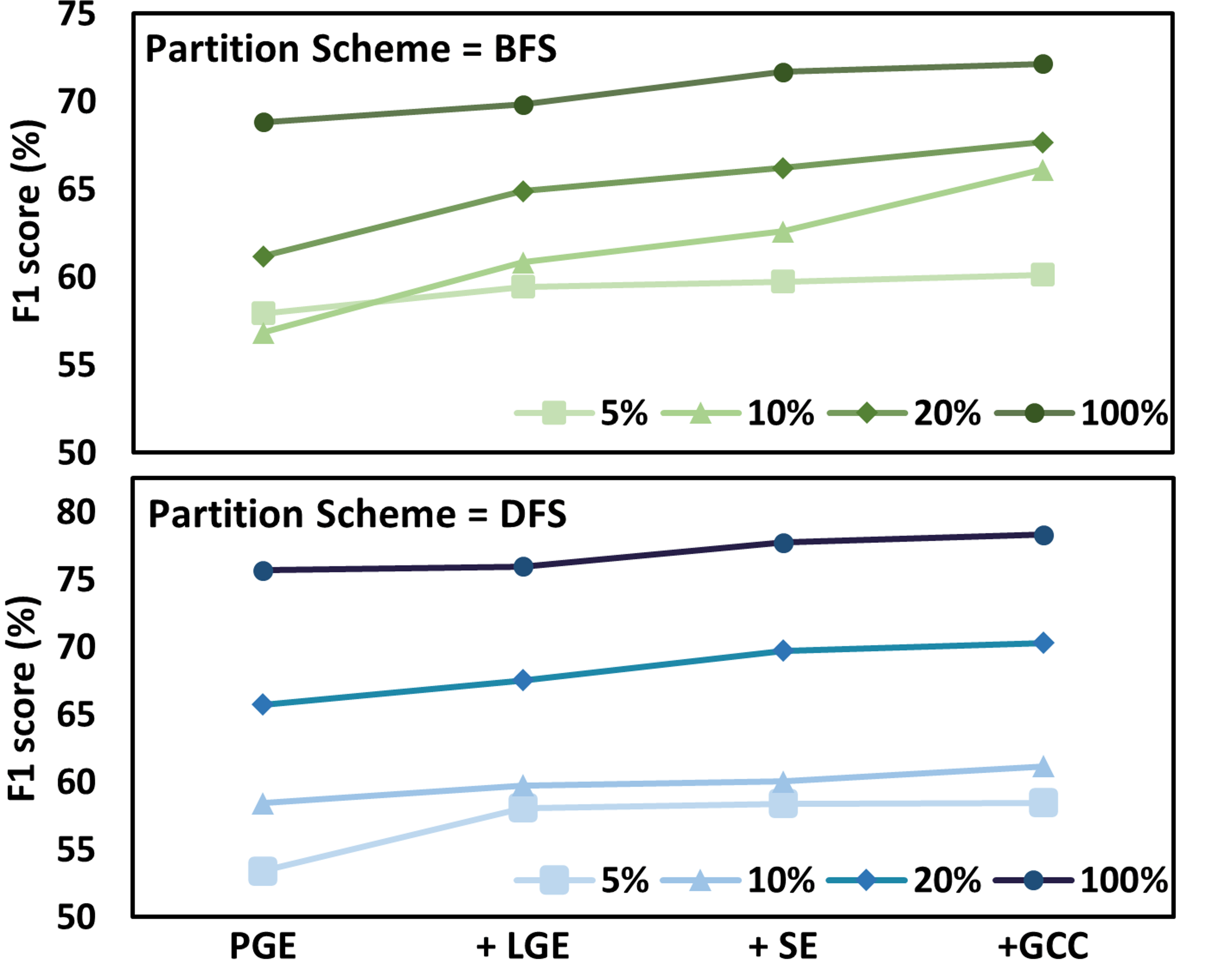}
    \caption{Results of ablation studies on different components of SemiGNN-PPI using the SHS27k dataset.}
    \label{fig:ablation}
\end{figure}

\noindent\textbf{Ablation Study.}
We investigate the effectiveness of different components in SemiGNN-PPI in Fig.~\ref{fig:ablation}. We can see that all components,~\emph{i.e.}, label graph encoding (LGE), self-ensemble (SE), and graph consistency constraint (GCC) positively contribute to the performance improvements. It is noted that too few labels~\emph{e.g.,} $10\%$ may influence model initialization, limiting self-ensemble graph learning, while the performance gains are more evenly distributed among various components with $10\%$ or more labeled data. Particularly, the proposed GCC can further enhance the results from the self-ensemble by providing stronger regularization in the feature space. Moreover, We
have performed one experiment for each augmentation strategy (F1-score) on SHS27k (20\% labeled) under random partition,~\emph{i.e.}, random edge dropout (78.92), random node dropout (79.04), centrality-based~\cite{tang2015cytonca} node and edge manipulation (81.08), and ours (81.11), which show that our strategy is comparable with centrality-based manipulation and outperforms others.

% It is noted that with $5\%$ labeled data, the performance increment is primarily from LGE rather than SE or GCC, in which the amount of labeled data is too little to effectively initialize models for self-ensemble learning. The performance gain is more evenly distributed among the proposed components with $10\%$ or more labeled data. Particularly, the proposed GCC further enhances the results from the self-ensemble by providing stronger regularization in the feature space.

% Comparing the results with different ratios of labeled data,

%Meanwhile, the overall performance increment is larger with the BFS partition scheme than with DFS for a fixed label percentage, as the testset created by BFS generally covers more PPIs between unseen proteins. 

\section{Conclusion}
In this paper, We propose a novel self-ensembling multi-graph neural network (SemiGNN-PPI) for efficient and generalizable multi-type PPI prediction, which models both protein correlations and label dependencies by constructing and processing graphs at protein and label levels. To leverage unlabeled PPI data, We integrate GNN into Mean Teacher for self-ensemble graph learning, in which multiple graph consistency constraints are designed to align the teacher and student graphs in the feature embedding space for optimized consistency regularization. Extensive experiments have demonstrated the superiority in model performance, label efficiency and generalization ability of SemiGNN-PPI over state-of-the-art methods by large margins.

\clearpage
\section*{Acknowledgements}
This research was funded by Competitive Research Programme ``NRF-CRP22-2019-0003", National Research Foundation Singapore, and partially supported by A*STAR core funding.

%  Disclaimer: The views expressed in this paper are
% solely those of the authors and do not necessarily represent
% the views of their current or previous employers.

\bibliographystyle{named}
\bibliography{ijcai23}

\end{document}